\documentclass[preprint]{aastex63}

\def\degree{\ifmmode^\circ\else$\null^\circ$\fi}
\def\degreespace{\ifmmode^\circ \else$\null^\circ$ \fi}
\usepackage[normalem]{ulem}
\usepackage{multirow}
\usepackage{graphicx}
\usepackage{array}
\usepackage{rotating}
\usepackage{amsmath}

\received{2021 June 17}
\revised{2021 August 5}
\accepted{2021 August 11}
\shorttitle{TPW + Infill}
\shortauthors{Johnson et al.}
\graphicspath{{./}{figures/}}

\begin{document}

\title{New Constraints on Pluto's Sputnik Planitia Ice Sheet from a Coupled Reorientation-Climate Model}

\correspondingauthor{Perianne E. Johnson}
\email{perianne.johnson@colorado.edu}

\author[0000-0001-6255-8526]{Perianne E. Johnson}
\affiliation{University of Colorado, Boulder}

\author[0000-0002-4803-5793]{James T. Keane}
\affiliation{Jet Propulsion Laboratory, California Institute of Technology}

\author{Leslie A. Young}
\affiliation{Southwest Research Institute}

\author{Isamu Matsuyama}
\affiliation{Lunar and Planetary Laboratory, University of Arizona}



\begin{abstract}
We present a coupled reorientation and climate model, to understand how true polar wander (TPW) and atmospheric condensation worked together to create the Sputnik Planitia (SP) ice sheet and reorient it to its present-day location on Pluto. SP is located at 18\degree N, 178\degree E, very close to the anti-Charon point, and it has been previously shown that this location can be explained by TPW reorientation of an impact basin as it fills with N$_2$ ice. We readdress that hypothesis while including a more accurate treatment of Pluto's climate and orbital obliquity cycle. Our model again finds that TPW is a viable mechanism for the formation and present-day location of SP. We find that the initial impact basin could have been located north of the present-day location, at latitudes between 35\degree N and 50\degree N. The empty basin is constrained to be 2.5 -- 3 km deep, with enough N$_2$ available to form at most a 1 -- 2 km thick ice sheet. Larger N$_2$ inventories reorient too close to the anti-Charon point. After reaching the final location, the ice sheet undergoes short periods of sublimation and re-condensation on the order of ten meters of ice, due Pluto's variable obliquity cycle, which drives short periods of reorientation of a few km. The obliquity cycle also has a role in the onset of infilling; some initial basin locations are only able to begin accumulating N$_2$ ice at certain points during the obliquity cycle. We also explore the sensitivity of the coupled model to albedo, initial obliquity, and Pluto's orbit.

\end{abstract}



\section{Introduction} \label{sec:intro}
Sputnik Planitia (SP), the western half of Pluto's ``heart,'' is a 1000 km-wide, several km-thick ice sheet made of nitrogen, methane, and carbon monoxide ices, located very close to the anti-Charon point on Pluto's surface \citep{Stern+2015}. The volatile ice sheet partially fills a topographic depression in  Pluto's water-ice crust. The ice sheet encompasses 870,000 km$^2$ ($\sim$ 4$\%$ of Pluto's surface area) and lie 2 km below the average surface elevation \citep{Schenk+2018}. Due to the topographic low and elliptical shape, the ice sheet is thought to lie within an ancient impact basin \citep{Stern+2015,Moore+2016,Johnson+2016}. Based on spectral modeling of the surface, the ice sheet is thought to be $\sim$ 50\% N$_2$ ice \citep{Protopapa+2017}, but the relative amounts of the various other ices (CH$_4$, CO) at depth is not well known. The presence of water ice blocks that appear to be floating on the edge of the ice sheet suggests that N$_2$ is the dominant species at depth, since methane's low density ($\sim$ half that of water ice) would not allow water ice blocks to float \citep{McKinnon+2016}. 

While SP contains a large amount of Pluto's N$_2$, there are also smaller ice deposits elsewhere on the surface, and the atmosphere contains some N$_2$ as well. \citet{GleinWaite2018} estimate that in Pluto's present-day atmosphere, the surface N$_2$ ice deposits contain five orders of magnitude more N$_2$ than the atmosphere. The amount of surface N$_2$ ice outside of SP is hard to estimate, because the depths of the observed deposits are unknown. Modelling by \citet{Bertrand+2019} found that perennial deposits of N$_2$ could be around 1 - 10 m thick. Additionally, the non-SP deposits must be thick enough to be observed spectroscopically. SP covers 870,000 km$^2$, and New Horizons observed another 6,200,000 km$^2$ to be N$_2$-covered. Since SP is estimated to be a few km thick, if the average thickness of the non-SP deposits is less than 0.5 km, then SP would be the dominant source of N$_2$, volumetrically. 

The surface of the ice sheet is characterized by a polygonal pattern, thought to be evidence of active solid-state convection \citep{Moore+2016}. The cells are a few tens of km in diameter, and the centers of the cells rise as much as 50 m above their edges. Convection modeling has shown that nitrogen ice sheets thicker than 1 km should undergo convection on Pluto, and that ice thicknesses of 3-6 km are necessary to explain the cell diameters, although the uncertain rheology of nitrogen ice introduces large uncertainties \citep{McKinnon+2016}. The convection models of \citet{Trowbridge+2016} calculate a slightly larger thickness of 10 km. \citet{McKinnon+2016} estimate that the basin itself should not be deeper than $\sim$10 km. 

Based on the lack of observed craters, an upper limit of 30 - 50 My is estimated for the crater-retention age of SP's surface \citep{Singer+2021}. However, the convective overturning motions of the ice sheet could refresh the surface in only 500,000 years \citep{McKinnon+2016, Buhler+2018}, so the ice sheet's surface is very young. Conversely, the age of the basin itself is very old, likely older than 4 Gy \citep{Singer+2021}, indicating that the infilling of the impact basin with N$_2$ ice could have occurred very early in Pluto's history. We show in this work that modeled basins collect all available N$_2$ very quickly, within 10 My typically, so infill likely occurred very soon after the impact, provided that there was sufficient available mobile N$_2$ on the surface. 

SP is centered around 18\degree N, 178\degree E, which places it very close to the anti-Charon point. \citet{Nimmo+2016} calculate that there is only a 5$\%$ chance of the basin forming that close to the anti-Charon point, and instead suggest that the basin likely migrated there as a result of true polar wander (TPW), a hypothesis shared by \citet{Keane+2016} as well. For this hypothesis to be true, the impact basin and ice sheet combination needs to be a positive gravity anomaly, since positive gravity anomalies reorient towards the equator to maintain a minimum energy state of principal axis rotation \citep{Matsuyama+2014}. Negative gravity anomalies, which one might expect an empty impact basin to be, reorient to one of the rotation poles for the same reason. However, large impact basins throughout the solar system are not predominantly negative gravity anomalies, but instead can be positive or negative anomalies based on various factors \citep{Keane+2016}. 

\citet{Keane+2016} find that a N$_2$ ice sheet alone is not able to explain SP's present-day location; an unreasonable amount of ice (enough to form a 100+ km-thick ice sheet) would be required to increase the empty basin's gravity anomaly enough to reorient it to the present-day location. \citet{Nimmo+2016} reach a similar conclusion, that the present-day location cannot be explained by TPW from N$_2$ infilling alone. Instead, the authors suggest that an ejecta blanket from the impact or subsequent uplift from a subsurface liquid ocean, or a combination of both, contribute to the positive gravity anomaly of the basin, in addition to the ice sheet. 

\citet{Keane+2016} compared the observed faults on Pluto to modeled faults resulting from reorientation stresses, and found that their large reorientation solutions (in which the impact basin starts northwest of SP's present-day location) produced faults most consistent with the observed pattern. The small true polar wander solutions (in which the impact basin starts north of SP's present-day location) were also consistent, but less so. This consistency strengthens the argument that TPW is responsible for the present-day location of SP.

\begin{figure}
	\begin{center}
		\includegraphics[width =\textwidth]{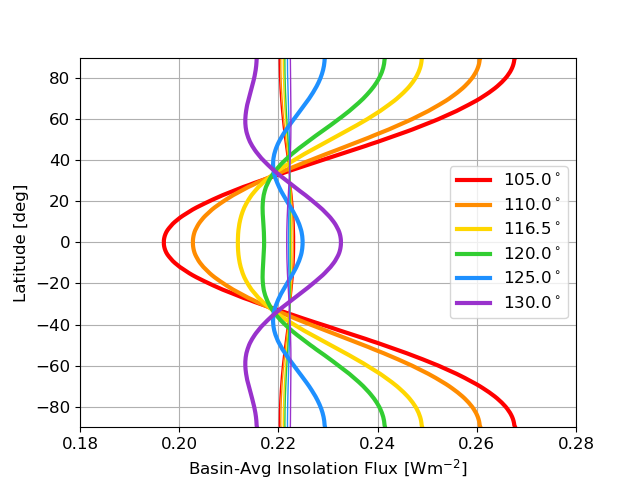}
		\caption{Orbit-averaged insolation onto a 24\degree-radius circular basin centered at each latitude for 6 example obliquities, chosen to span Pluto's obliquity cycle. Pluto's current obliquity is 120\degreespace (green curve) and increasing towards the maximum obliquity of 128\degree. The thin lines with little variation represent the spatially-averaged insolation onto the rest of the surface, excluding the basin, for each obliquity. If the effect of depth is ignored, a basin will fill with N$_2$ ice when the basin-averaged insolation is less than the surface-averaged insolation.}
		\label{fig:insol}
	\end{center}
\end{figure}

\citet{Hamilton+2016} suggest an alternative to the impact basin formation theory for SP. They suggest that N$_2$ ice will naturally accumulate at latitudes near 30\degree, because, when averaged over Pluto's orbit, that region is the coldest part of the surface. The runaway albedo affect can concentrate the ice into a single cap at 30\degree N within 1 My after the Charon-forming impact, and the subsequent orbital evolution locks the mass anomaly of this cap onto the anti-Charon side, as is observed. Finally, the weight of the many km-thick accumulation of ice causes the underlying crust to slump, creating the observed topographic depression. This hypothesis requires that latitudes near SP's present-day latitude are the coldest region on the surface, which is true for some, but not all orbital obliquities in Pluto's present-day obliquity cycle, as shown in Figure \ref{fig:insol}. \citet{Hamilton+2016} did not consider the expected global tectonic pattern in this scenario. The quasi-radial fractures proximal to SP can be explained by N$_2$ loading \citep{Keane+2016,McGovern+2021}. However, the quasi-azimuthal faults distal to SP require TPW stresses \citep{Keane+2016}.

Pluto's obliquity (the angle between the spin axis and the heliocentric orbital axis) varies over time due to perturbations from the rest of the solar system bodies \citep{DobrovolskisHarris1983}. \citet{Dobrovolskis+1997} calculated that the obliquity would vary between 103\degreespace and 128\degreespace over a 2.7 My period. Pluto's spin is retrograde, so the obliquity value is reported as $>$90\degree, and higher values indicate smaller absolute angles between the spin and orbit axes (less ``tilting''). Pluto's obliquity is currently 120\degreespace and increasing with time. The obliquity of the orbit affects the latitudinal distribution of incident insolation, as shown in Figure \ref{fig:insol}. To speedily calculate the orbital-averaged insolation, we implement the sixth order Legendre approximation from \citet{NadeauMcGehee2017}. Figure \ref{fig:insol} shows the orbit-averaged incident insolation onto the surface, averaged over the area inside the modeled 24\degree-radius basin. For the local (not basin-averaged) insolation, see Fig. 1 in \citet{Hamilton+2016}, for example. Averaging over the area of the basin has the effect of flattening the insolation curves, and we perform this averaging because the model used here calculates the average infill rate into the basin, which depends on the average insolation. Higher obliquity values (indicating less tilted orbits) have a more uniform distribution (e.g. the purple curve), while lower values exhibit extreme contrasts between the annual average insolation at the poles vs. at the equator (e.g. the red curve). Thus, the obliquity cycle has significant consequences for an impact basin's ability to fill in with N$_2$ ice from atmospheric condensation. At a time of low obliquity, a high-latitude basin might receive more insolation than the rest of the surface, so condensation does not occur, but sometime later in the obliquity cycle the basin will receive equal or less insolation than the rest of the surface. Lower insolation, or equal insolation coupled with the depth of the basin, leads to condensation of N$_2$ ice onto the basin floor.

The model presented here couples TPW with climate for the first time, to understand if, and how, these processes interact to place SP at its present-day location. This work seeks to determine for which initial conditions (basin depth, N$_2$ inventory, and basin location) is the present-day location of SP consistent with infill of an impact basin via atmospheric condensation (accounting for a cyclic obliquity) and subsequent reorientation via TPW. The methodology is explained in Section \ref{sec:model}, and the infill and reorientation results for a wide variety of initial locations and basin depths are shown in Section \ref{sec:results}. Sections \ref{sec:obliquity} -- \ref{sec:sma} contain more details about the effect of initial obliquity, crustal deformation, albedo, and orbital semi-major axis on the results. The conclusions are summarized in Section \ref{sec:conclusions}.


\section{Model} \label{sec:model}

\subsection{Volatile Infill Model}

The model is initialized in a ``snowball'' Pluto state, in which the available N$_2$ ice is spread uniformly over the entire surface. The model tracks the average thickness of N$_2$ ice in two regions: inside of a 24\degree-radius circular basin (subscript ``basin'') and outside of this basin on the rest of Pluto's surface (subscript ``surface''). It is assumed that any ice that sublimes from one region immediately condenses uniformly over the other region. This assumption of instantaneous mixing of nitrogen in the atmosphere was also used by \citet{Bertrand2016}. We do not account for a variable atmospheric mass, although we do assume that there is always sufficient N$_2$ mass in the atmosphere to maintain a global atmosphere in which temperature is independent of latitude and longitude (but temperature does depend on altitude). Modeling by \citet{Johnson+2021} showed that Pluto's atmosphere remains global over the past 3 My for most realistic scenarios. The two regions have equivalent Bond albedos of 0.75, and the albedo is constant in time throughout the model run, so we do not account for an albedo that varies based on the abundance of N$_2$ on the surface, for example. We explore the effect of different albedos in Section \ref{sec:albedo}.

The infilling rate $\dot{m}$ (in kg s$^{-1}$ m$^{-2}$) is calculated by assuming energy balance and global mass conservation between the two regions. In each region, the difference between absorbed insolation and thermal emission gives the latent heat energy of the condensed material:

\begin{equation}
    \Omega_{surface}\epsilon\sigma T_{surface}^4 - \Omega_{surface}(1-A)S_{surface} = \Omega_{surface} L \dot{m} _{surface} 
\end{equation}
\begin{equation}
    \Omega_{basin}\epsilon\sigma T_{basin}^4 - \Omega_{basin}(1-A)S_{basin} = \Omega_{basin} L \dot{m} _{basin} 
\end{equation}

$\Omega$ is the angular area of the region, $\epsilon$ is the emissivity, $\sigma$ is the Stefan-Boltzmann constant, $T$ is the N$_2$ ice temperature, $S$ is the absorbed insolation (averaged over the region and averaged over one Pluto orbit), and $L$=2.5 x 10$^5$ J kg$^{-1}$ is the latent heat of N$_2$. We have not cancelled the $\Omega$ factors in Equations 1 and 2 to make it clear that $\dot{m}$ is a per-area quantity. 

Global mass balance is also enforced, requiring that any N$_2$ sublimed from one region condenses onto the other region:
\begin{equation}
    \Omega_{surface} \dot{m}_{surface} = -\Omega_{basin} \dot{m}_{basin}
\end{equation}
In reality, some of the sublimed N$_2$ stays in the atmosphere, rather than immediately condensing onto the other region. However, the atmospheric mass is several orders of magnitude smaller than the mass of N$_2$ on the surface, so these seasonal variations in atmospheric mass are negligible to the total N$_2$ inventory \citep{GleinWaite2018}. For example, the atmospheric pressure above a N$_2$-covered surface at 34.3 K (the surface temperature for most of our modeled scenarios) is about 1.8 $\mu$bar, which implies an atmospheric mass of 5 x 10$^{12}$ kg. 40 - 80 m GEL, the range of N$_2$ inventories we consider here, is equivalent to 0.7 - 1.4 x 10$^{18}$ kg of N$_2$ ice.

To arrive at an equation for the infilling rate, we need to relate the two N$_2$ ice temperatures. The pressure at the top of the ice sheet is larger than at the non-basin surface because of hydrostatic equilibrium, and the equilibrium temperature is consequently higher at the top of the ice sheet via the Clausius-Clapeyron relation. We have defined the ratio of the temperature at the top of the ice sheet $T_{basin}/T_{surface}$ by a factor $a$. If we assume that the atmospheric temperature is in equilibrium with the pressure at every at altitude (that is, the gradient follows the wet adiabat of the primary gas), then
\begin{equation}
    a = e^{gd/L}
\end{equation}
where $d$ is the depth to the top of the ice sheet and $g$ = 0.62 ms$^{-2}$ is Pluto's gravitational acceleration. For $T_{basin}$=37 K and $d$=3 km,  $a$=1.0075.
Since the depths are small compared to a scale height, $a$ depends only slightly on the thermal structure of the atmosphere above the top of the ice sheet. For example, if the atmosphere were isothermal above the ice sheet, then $a = 1+gd/L$, which alters $T_{basin}$ by of order 1 mK.


Combining equations 1 through 4 leads to the following equation for the rate of infill into the basin:

\begin{equation}
    \dot{m}_{basin} = (1-A) \frac{\Omega_{surface}}{\Omega_{surface} + a^4\Omega_{basin}}\left(a^4S_{surface} - S_{basin}\right) \frac{1}{L}
\end{equation}

At each timestep, the model calculates the average insolation onto each region and uses the current ice sheet depth to calculate the rate of condensation onto (or sublimation from) the top of the ice sheet. This added ice mass is used to update the gravity anomaly of the basin and calculate the corresponding reorientation of Pluto. 

\subsection{True Polar Wander (TPW) Model}

The TPW part of the model closely follows the method used in \citet{Keane+2016}, which we summarize here. We created a simple dynamical model of Pluto, where the total inertia tensor of Pluto, $\mathbf{I}_{Pluto}$, is written as the sum of two components: the remnant figure, $\mathbf{I}_{RF}$, and the contribution from SP, $\mathbf{I}_{SP}$:
\begin{equation}
    \mathbf{I}_{Pluto} = \mathbf{I}_{RF} +  \mathbf{I}_{SP}
\end{equation}
where bold symbols indicate tensors. Only the non-spherically symmetric contribution to the inertia tensors plays a role in determining the amount of TPW because the spherically symmetric contribution does not modify the principal axes of inertia. We take into account deformation of Pluto in response to both the changing tidal/rotational potential and response due to volatile loading of SP using Love number theory \citep[e.g.][]{Sabadini+2016}. Given the long $\sim$Myr timescales involved, we use the long-term Love numbers describing long-term deformations.  

The remnant figure, $\mathbf{I}_{RF}$, represents Pluto's non-hydrostatic, elastically supported tidal-rotational bulge. This bulge is preserved no matter how Pluto reorients. At present, no bulge has been observed at Pluto, although \citet{Nimmo+2017} report an upper limit of 0.6\% for Pluto's oblateness. Thus, we construct a theoretical remnant figure using Love numbers  \citep[e.g.][]{Matsuyama+2014}. We assume the nominal four-layer interior structure of Pluto from \citet{Keane+2016} consisting of a silicate-rich core, liquid water ocean, and a two-layer water ice shell with a weak lower crust and upper, 50 km thick elastic lithosphere with a rigidity of 3.5 GPa. The remnant figure arises due to deformation in response to rotation and the present-day tides of Charon, and is supported entirely in the elastic lithosphere. The remnant figure could relax over time, but the timescale for this relaxation is unknown. If we consider a thinner elastic lithosphere, there will be more compensation to the ice load, which leads to smaller TPW reorientations. However, a thinner lithosphere also leads to a smaller remnant figure, which has a stabilizing effect and makes TPW reorientations larger. Therefore, the relaxation of the remnant figure may not have a significant effect on the results presented here. A detailed study of the effect that the magnitude of the remnant figure has on SP's initial location and the ice thickness is beyond the scope of this work, so we only use our nominal value of 50 km for the elastic thickness. 

The inertia tensor from SP, $\mathbf{I}_{SP}$, represents the total inertia tensor perturbation arising from SP, which we subdivide further into the inertia tensor arising from the underlying impact basin and the ice sheet:
\begin{equation}
    \mathbf{I}_{SP} = \mathbf{I}_{basin} +  \mathbf{I}_{ice \: sheet}
\end{equation}
The inertia tensor from the underlying basin is unknown. In the absence of data, we agnostically assume that the basin was initially fully compensated ($\mathbf{I}_{basin}=0$), which is a reasonable assumption if SP has an underlying ocean uplift or surrounding ejecta blanket \citep{Keane+2016,Nimmo+2016}. With this assumption, the only contribution to $\mathbf{I}_{SP}$ is from the ice sheet, which we describe as a spherical cap of uniform thickness. For the case of SP located on the north pole of Pluto, the uncompensated ice sheet contribution can be written: 

\begin{equation}\label{inertia_tensor_ice_sheet_unc}
    \mathbf{I}_{ice \: sheet}^\prime = 
     \frac{\pi \sigma r^4 (4- 3\cos \gamma)}{3}  \begin{bmatrix}
  1  & 0 & 0 \\
    0 & 1  & 0 \\
    0 & 0 & 1
    \end{bmatrix}
  +\frac{\pi \sigma r^4 \cos \gamma}{3}  \begin{bmatrix}
   -\cos^2 \gamma  & 0 & 0 \\
    0 & -\cos^2 \gamma  & 0 \\
    0 & 0 & 2 \cos^2 \gamma -3
    \end{bmatrix}
\end{equation}
where $\sigma$ is the surface density (mass per unit area, i.e., kg m$^{-2}$) of the ice sheet, $\gamma$ is the angular radius of the ice sheet, and $r$ is the radius of Pluto (1187 km). $\mathbf{I}_{ice \: sheet}^\prime=0$ when $\gamma=0$, as expected.
The first term in Eq. (\ref{inertia_tensor_ice_sheet_unc}) does not play a role in inducing TPW because it is spherically symmetric, and therefore it is ignored hereafter.
Taking compensation into account, 
the non-symmetric ice sheet contribution can be written: 

\begin{equation}\label{inertia_tensor_ice_sheet}
    \mathbf{I}_{ice \: sheet} = 
  (1+k_2^L)\frac{\pi \sigma r^4 \cos \gamma}{3}  \begin{bmatrix}
   -\cos^2 \gamma  & 0 & 0 \\
    0 & -\cos^2 \gamma  & 0 \\
    0 & 0 & 2 \cos^2 \gamma -3
    \end{bmatrix}
\end{equation}
where $k_2^L$ is a degree two load Love number describing the long-term deformation in response to ice loading.  Without an elastic lithosphere, the ice sheet would be fully compensated ($k_2^L=-1$) and would not contribute to the inertia tensor. Conversely, in the limit case of an infinite rigidity elastic lithosphere, Pluto would not deform in response to ice loading ($k_2^L=0$). For the assumed nominal four-layer  interior structure containing a 50 km thick elastic lithosphere with a rigidity of 3.5 GPa, $k_2^L=-0.55$. While Eq. (\ref{inertia_tensor_ice_sheet}) describes the case for the ice sheet located on the north pole, it can be rotated to anywhere on the globe using standard rotation matrices. 

\subsection{The Combined Model}

The reorientation is calculated iteratively. We start first with Pluto's remnant figure, and consider an arbitrary initial location, and in each subsequent timestep, we calculate how much mass is added to (or lost from) the ice sheet. After each step, we diagonalize the resulting inertia tensor, and define the new rotation axis and tidal axis as the maximum and minimum principal axes of inertia, respectively. This is equivalent to keeping Pluto in a minimum energy (principal axis) rotation state---which implicitly assumes that Pluto can rapidly adjust to applied forces. The angles between the new and old principal axes of inertia give the reorientation of the surface and determine the new latitude and longitude of the basin and ice sheet.

As the ice sheet grows in mass, the underlying basin and surrounding area deforms, with the basin floor sinking in elevation and the area immediately surrounding the basin rising slightly. We assume a spherical cap with an angular radius of 24\degreespace for the ice sheet and compute the corresponding radial deformation using displacement Love numbers.  
The deformation is larger near the basin center with an amplitude and profile similar to the one found by \citet{McGovern+2021} using a finite element method. In order to find the average deformation, we convert the modelled bowl-shaped deformation profile to a flat-bottomed, cylindrical profile. We do this by finding the volume of ice contained within the 24\degree-radius deformed ice sheet and equating it to the volume of a flat-bottomed, cylindrical idealized ice sheet of some thickness. The bottom elevation of the ice sheet implied by this thickness gives the average deformation. For every 1 km of N$_2$ ice in the basin, the basin floor sinks by roughly 0.5 km, comparable to the average deformation in \citet{McGovern+2021} (sinking an average of 1.27 km for every 2.27 km of N$_2$ ice, a deformation-to-load ratio of 0.56 to our ratio of 0.5). This affects the model in two ways: it reduces the contribution of the ice sheet load to the overall inertia tensor, and it increases the infill rate by lowering the elevation of the condensation surface at the top of the ice sheet. Section \ref{sec:deformation} explores the effect of ignoring basin floor deformation. 

As Pluto reorients, the basin and nascent ice sheet are moved to different latitudes and therefore receive different amounts of average insolation. The model accounts for Pluto's varying obliquity, which also varies the insolation onto the basin as a function of time. Pluto's obliquity currently varies on a 2.7 My cycle, between 105\degreespace and 129\degreespace \citep{Earle+2017}. It is not clear what Pluto's obliquity would have been at the time of the impact that formed the basin and/or at the time of the infilling, but the model assumes the current-day obliquity cycle. 

For this work, we investigated initial basin latitudes between 5\degreespace and 85\degreespace latitude and 95\degreespace and 175\degreespace longitude, in increments of 5\degreespace (the initial location of SP is limited to this single octant of the surface). The model timestep was 300 Pluto orbits, about 75,000 years. The reorientation and the glacial flow needed to level the surface of the ice sheet were both assumed to happen within a single timestep. The models ran for 268 timesteps, giving a total run time of 20 My, at which point the modelled ice sheet had reached a semi-stable final location and thickness. The basin was circular in cross section, flat-bottomed, and covered 4\% of Pluto's surface, the equivalent of present-day SP's extent. We used Pluto's current orbital parameters in the model, although alternative orbits, such as a smaller semi-axis orbit, are discussed in Section \ref{sec:sma}.

The top of SP is 2 km below the surface's average elevation \citep{Schenk+2018}. Thus, we only considered initial basin depth and N$_2$ inventory combinations that result in this configuration. In the following section, we present the results from initial basin depths of 2.5 and 3 km, paired with final ice thicknesses of 1 and 2 km, respectively. These correspond to N$_2$ inventories of 0.7 x 10$^{18}$ kg and 1.4 x 10$^{18}$ kg, and global equivalent layers (GEL) of 40 - 80 m. Deeper initial basins require larger N$_2$ inventories, which cause too much reorientation and place the basin too close to the sub-Charon point. 


\section{Results} \label{sec:results}
Runs are considered successful if the basin's final semi-stable location lies within $\pm$5\degreespace of the present-day center of SP. SP's center can be defined in several ways: the centroid of the ice sheet as defined from geologic maps \citep{White+2017}, the centroid of the larger topographic low that the ice sheet lies within, or the averaged latitude and longitude of the ice sheet region weighted by the surface area. We adopt an average of these three values as the center of SP: 17.7\degree N, 178.2\degree E. The model runs for 20 My, which is long enough for all cases to reach a final semi-stable state. The model ends at Pluto's present-day obliquity value of 120\degree. We only explore initial locations within one octant of Pluto's surface (the northeastern part of the western hemisphere: 0\degree N - 90\degree N, 90\degree E - 180\degree E), because TPW will not cause basins to cross the equator or the 90\degree E and 180\degree E longitudes. Since the present-day location of SP is within this octant, the initial location must be as well. 

\begin{figure}
	\begin{center}
		\includegraphics[width =\textwidth]{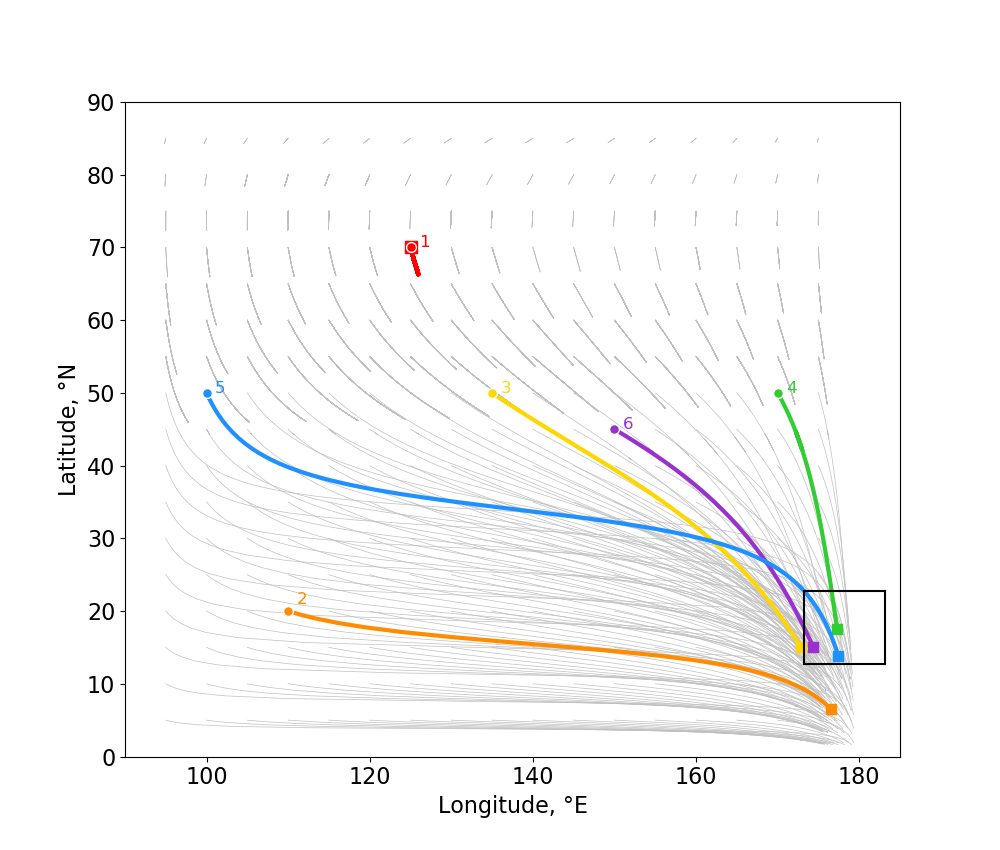}
		\caption{Reorientation paths for 3-km deep basins at all initial locations with 80 m GEL of available N$_2$ (faded lines). The solid black box marks the region within $\pm$5\degreespace of the present-day center of SP. A few example paths are highlighted: (1, red) an initially high-latitude basin's path; (2, orange) an initially low-latitude basin's path; (3, yellow) the path of an initially intermediate-latitude basin that ends too far west; (4, green) the path of an initially intermediate-latitude and high-longitude basin that ends within $\pm$5\degreespace of the present-day center of SP; (5, blue) the path of an initially intermediate-latitude and low-longitude basin that ends within $\pm$5\degreespace of the present-day center of SP; (6, purple) the path of an initially intermediate-latitude and intermediate-longitude basin that ends within $\pm$5\degreespace of the present-day center of SP. Small circles denote the initial location of the basin and small squares denote the final location. Note that for path 1, the initial and final locations are the same. The color scheme in subsequent figures (excluding Figure \ref{fig:success}) corresponds to the labeled paths in this figure.}
		\label{fig:paths}
	\end{center}
\end{figure}

Figure \ref{fig:paths} shows the paths that basins take in an instantaneous reference frame defined by the current timestep's principal axes, in which the latitude and longitude of surface features change over time. As an example, Figure \ref{fig:paths} shows the paths resulting from a 3 km deep basin with an available N$_2$ inventory of 80 m GEL. 

Basins that begin near the pole, at latitudes $\gtrsim $50\degree, never infill completely. An example of the corresponding reorientation is shown by the red line labeled as 1 in Figure \ref{fig:paths}. For most obliquity values, the poles receive higher insolation than lower latitudes, so ice sheets near the pole receive more sunlight than the N$_2$ deposits elsewhere on the surface and tend to sublime away. However, at certain points during the obliquity cycle, the obliquity is high enough, that when combined with the depth of the basin, the infill rate becomes positive and some condensation occurs. As a result, the basin reorients slightly, moving down in latitude. However, when the obliquity increases again, all of the condensed ice sublimes away and the basin reorients back to its initial location.

Basins that start near the equator, at latitudes $\lesssim $35\degree, collect all of the available N$_2$ inventory, but subsequently reorient too close to the sub-Charon point instead of stopping at a semi-stable final location close to the present-day center of SP. An example reorientation path is shown by the orange line labeled as 2 in Figure \ref{fig:paths}. Similarly, basins that begin at intermediate latitudes and longitudes, approximately 130\degree E to 150\degree E and 45\degree N to 50\degree N (see Region 3 in Figure \ref{fig:summary}), infill completely and collect all available N$_2$, but the resulting reorientation leaves the final basin too far west of the anti-Charon longitude. An example reorientation path is shown by the yellow line labeled 3 in Figure \ref{fig:paths}. 

There are three regions of Pluto's surface that lead to successful reorientations. First, basins that begin at intermediate latitudes ($\sim$35\degreespace N - 50\degreespace N) directly north of the present-day location infill with all available N$_2$ and reorient to final locations within $\pm$5\degreespace of the present-day location of SP. The green line labeled 4 in Figure \ref{fig:paths} shows an example path, analogous to the ``small-TPW'' solutions described in \citet{Keane+2016}. Additionally, basins that begin at low initial longitudes and intermediate latitudes also reorient to final locations within $\pm$5\degreespace of the present-day center of SP. An example reorientation path is shown by the blue line labeled 5 in Figure \ref{fig:paths}, analogous to the ``large-TPW'' solutions described in \citet{Keane+2016}. Basins that begin at intermediate longitudes in a narrow range of intermediate latitudes (see Region 6 in Figure \ref{fig:summary}) reorient to final locations within $\pm$5\degreespace of the present-day center of SP. An example reorientation path is shown by the purple line labeled 6 in Figure \ref{fig:paths}.

\begin{sidewaysfigure}
	\begin{center}
		\includegraphics[width =\textwidth]{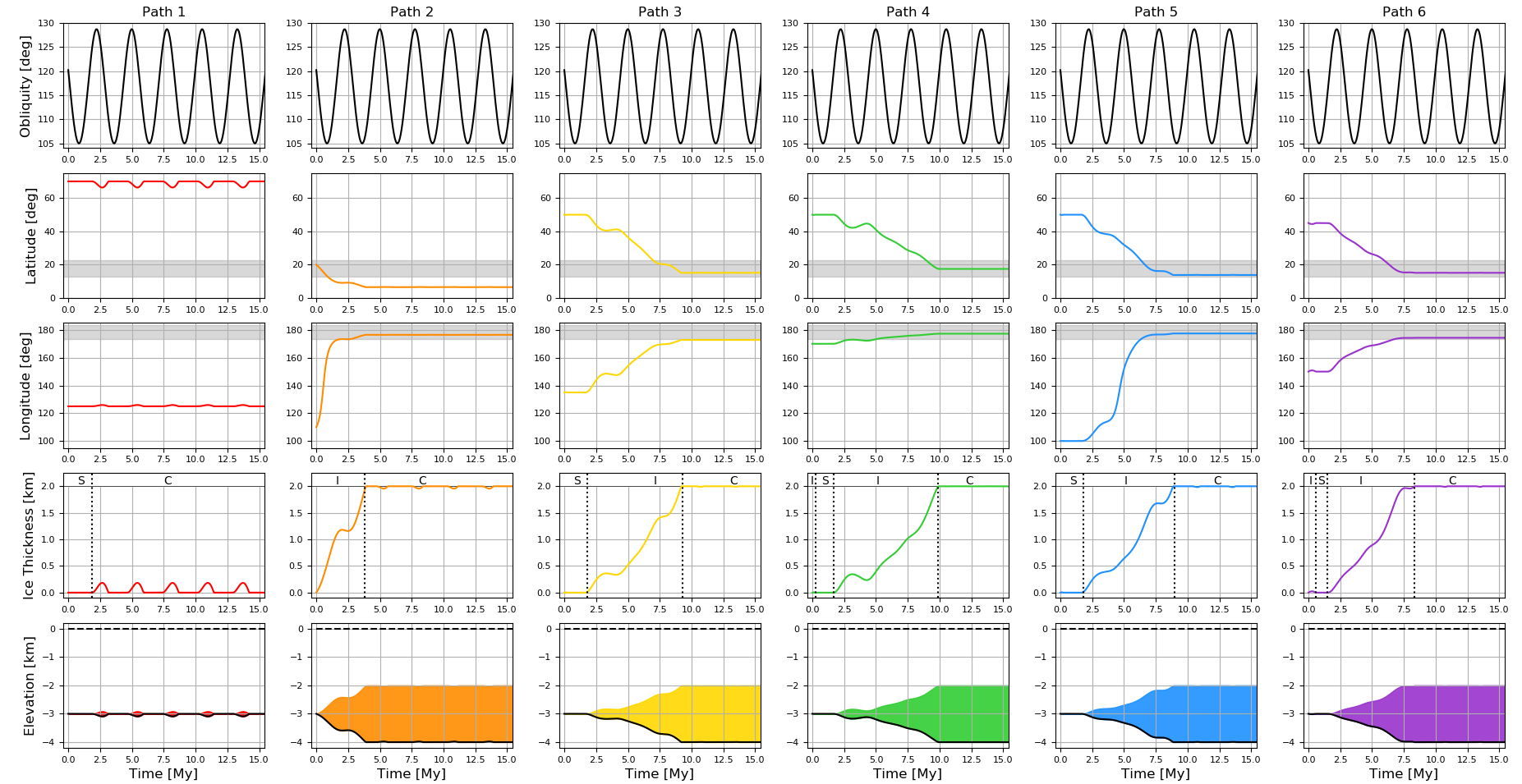}
		\caption{Temporal behavior for Pluto's obliquity, the basin's latitude and longitude, ice thickness, and the cross-section of the basin for each path. For the latitude and longitude rows, the shaded region is $\pm$5\degreespace from the latitude or longitude of SP's present-day location. Obliquity is shown in black because it is the same for all paths; it is repeated in each column for ease of comparison. In the ice thickness plots (second to bottom row), three phases of behavior are labeled: (S) Stationary, (I) Infilling, and (C) Cyclic. The bottom row shows a cross section of the basin in time, with the basin floor shown as a solid black line, the surface shown as a dotted black line at 0 km elevation, and the ice sheet shown as solid color fill. The color scheme is the same as that used in Figure \ref{fig:paths}.}
		\label{fig:bigfig}  
	\end{center}
\end{sidewaysfigure}

Figure \ref{fig:bigfig} shows the temporal behavior of Pluto's obliquity, the basin's location, and the ice thickness for each path drawn in Figure \ref{fig:paths}. The behavior can be classified into three phases: (S) Stationary, (I) Infilling, and (C) Cyclic, which are labelled in the fourth panel of Figure \ref{fig:bigfig}. A given basin may experience all three phases, or it may only experience one or two of the phases. The phases can be experienced in any order and can be experienced more than once. For example, Path 1 moves straight from the stationary phase to the cyclic phase, while Path 4 starts out in the infilling phase, briefly enters the stationary phase, and then returns to infilling before finally entering the cyclic phase. 

In the case of the blue path labeled 5 (initially intermediate latitude and low longitude basin), the basin first moves primarily south, then east, and then south again. The basin is stable at the initial location for $\sim$ 2 My (Stationary Phase), and then gradually gains ice mass and reorients over a period of about 6.5 My (Infilling Phase) once the obliquity reaches a high value and the insolation onto the basin is less than that onto the rest of the surface. When the obliquity is at high values (note that Pluto's obliquity is $>$90\degreespace due to its retrograde rotation, so a high obliquity value indicates a smaller angle between the spin pole and the orbit plane), the difference in insolation between the equator and poles is at a minimum. The basin remains semi-stable at the new location and ice thickness, only experiencing small deviations on the order of one tenth of a degree in latitude and longitude and tens of meters in ice thickness as the obliquity cycles (Cyclic Phase). These small excursions in location and ice thickness occur for basins at all initial locations, but vary in magnitude, with the initially high-latitude basins exhibiting the largest oscillations on the order of 5\degreespace latitude and 180 m in ice thickness. This is discussed more in Section \ref{sec:excursions}.

The preceding discussion used an initially 3-km deep basin with 80 m GEL available N$_2$ inventory (enough to create a 2 km thick ice sheet in the basin and a final basin depth of 4 km), but we explored other initial basin depths and N$_2$ inventory combinations as well. Combinations that create a final ice sheet at an elevation of $-2$ km are as follows: 2.5 km initial depth with a 1 km ice sheet, 3 km initial depth with a 2 km ice sheet, 3.5 km initial depth with a 3 km ice sheet, and so on. We found that ice sheets thicker than 2 km reoriented too close to the anti-Charon point for all initial basin locations that experienced an Infilling Phase. This result is subject to our assumption of the lithosphere's elastic thickness of 50 km. The effective size of a mass load inducing TPW is given by the ratio between the degree-2 gravity coefficients of the compensated load and the remnant figure \citep[][Eq. (20)]{Matsuyama+2014}. For a thicker elastic lithosphere, the compensated load increases, increasing TPW, and the remnant figure increases, decreasing TPW. These two effects balance each other, resulting in a weak dependence on the assumed elastic lithosphere thickness \citep[e.g.][Extended Data Figure 5k]{Keane+2016}.

\begin{figure*}
	\begin{center}
		\includegraphics[width =\textwidth]{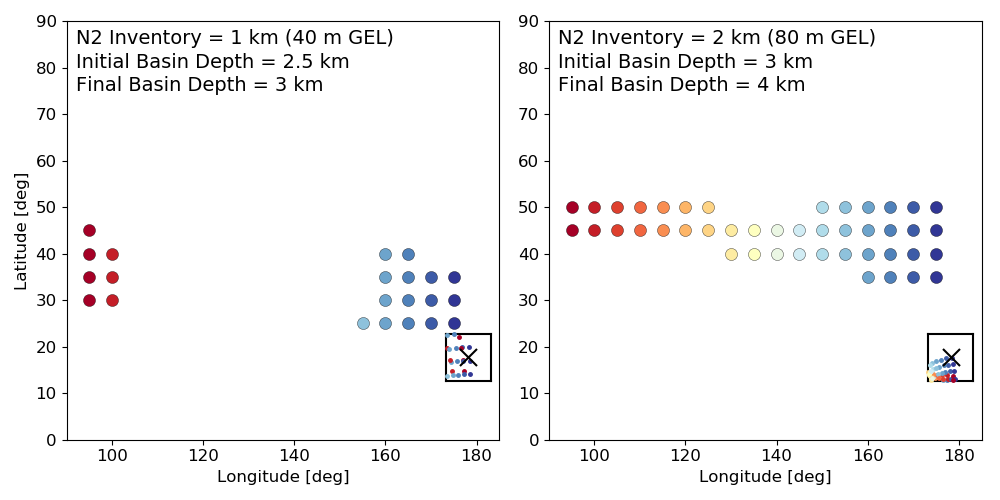}
		\caption{Initial locations for the impact basin that reorient to place the basin within 5 degrees of SP's current location are shown as small circles. The left panel shows the case for an initially 2.5 km deep basin and the right panel shows the case of an initially 3 km deep basin. The black cross and surrounding box show the present-day center of SP and the region $\pm$5\degreespace from the center. The final locations are also shown as small circles, and they tend to cluster in the southwest corner of the boxed success region. The initial longitude of the basins are shown via color: the western basins are red, intermediate-longitude basins are white, and eastern basins are blue.}
		\label{fig:success}
	\end{center}
\end{figure*}

\begin{figure}
	\begin{center}
		\includegraphics[width =\textwidth]{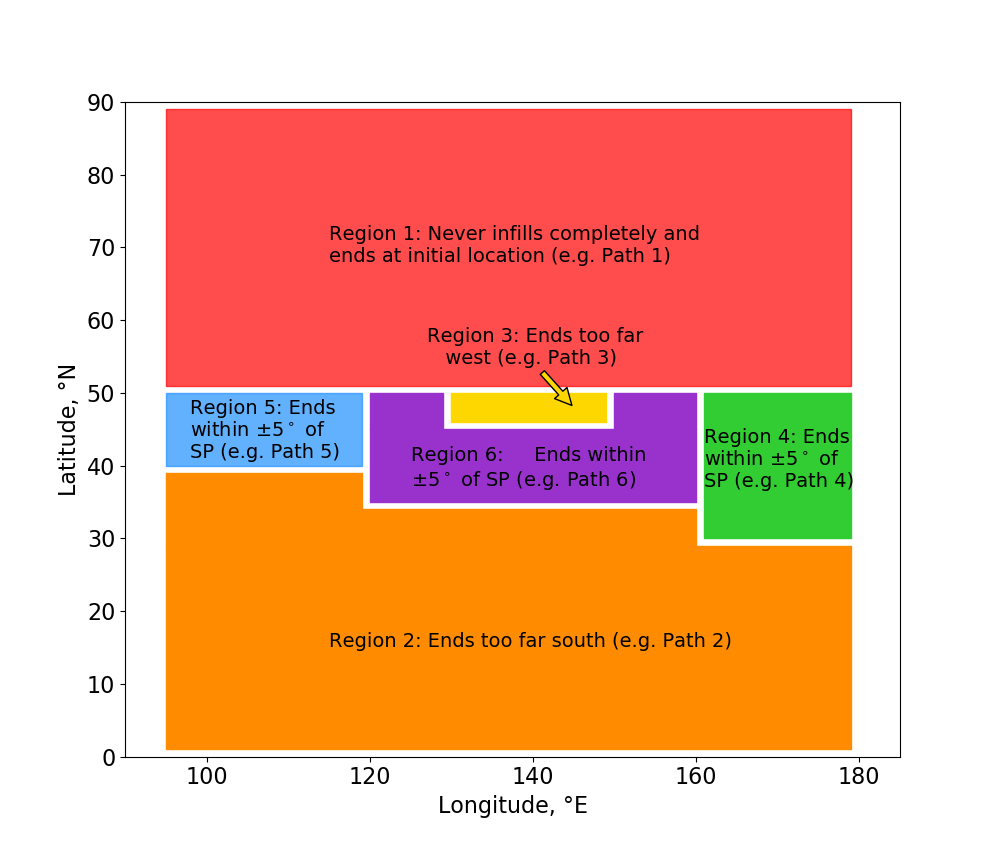}
		\caption{Graphical summary of the final location of the basin as a function of initial latitude and longitude, using the same color scheme as Figure \ref{fig:paths}. Initial basins that begin within each colored region will have final locations as described by the text within the region. }
		\label{fig:summary}
	\end{center}
\end{figure}

Figure \ref{fig:success} shows those initial basin locations for each initial basin depth that reoriented so that the final basin location was within $\pm$5\degreespace of the present-day location of SP, for the 2.5 and 3 km initial depth basins. The initial longitude of the basin is denoted with color in Figure \ref{fig:success}, with the color transitioning from red to white to blue as longitude increases. For the 3 km initial depth (right panel), basins that start far to the west of SP's present-day location (red circles) undergo large reorientations and tend to end up south of SP's present-day location (cluster in the bottom of the success box). Basins that start at intermediate longitudes (pale red to pale blue colors) end up west or southwest of SP's present day location, and basins that start directly north (blue circles) tend to have final locations that are the closest of the present-day location of SP, but there is significant spread in the final location. The initially 2.5 km deep basins have even more significant spread in their final locations. 


Figure \ref{fig:summary} summarizes the final location of the basin as a function of initial latitude and longitude. The color scheme is the same as Figure \ref{fig:paths}. The latitude and longitude boundaries of the regions are approximate, in order to be representative of various initial basin depths.

\subsection{Obliquity Cycle Excursions}\label{sec:excursions}
After some time in either the Stationary or Infilling Phases (or both), usually within 10 My, basins from all initial locations reach a semi-stable final location and ice thickness. However, they still experience short, small excursions from this location and thickness at certain parts of the obliquity cycle. \citet{Keane+2016} considered ``seasonal wobbles'' owing to the transport of volatiles on annual timescales; the obliquity cycle excursions discussed here are the analogous ``superseasonal wobbles.''
If the depth to the top of the ice sheet is zero (e.g. ice thickness equals basin depth), then the infill rate is proportional to the difference between the average insolation onto the basin/ice sheet and the average insolation onto the rest of the surface; if the basin receives more insolation, N$_2$ ice will sublime away from the ice sheet while if the basin receives less insolation, N$_2$ will condense onto the ice sheet. ice sheets set inside deep basins can receive slightly more insolation than the average of the rest of the surface (up to $e^{4gd/L}S_{surface}$) and still experience condensation. The relative insolation depends on the latitude of the basin and also on the current obliquity (since that controls the latitudinal distribution of insolation, see Figure \ref{fig:insol}). Even when a basin has reached a semi-stable location and ice thickness, the obliquity cycle causes variations in the relative insolation onto the basin and thus small excursions in latitude, longitude, and ice thickness occur.

\begin{figure*}
	\begin{center}
		\includegraphics[width =\textwidth]{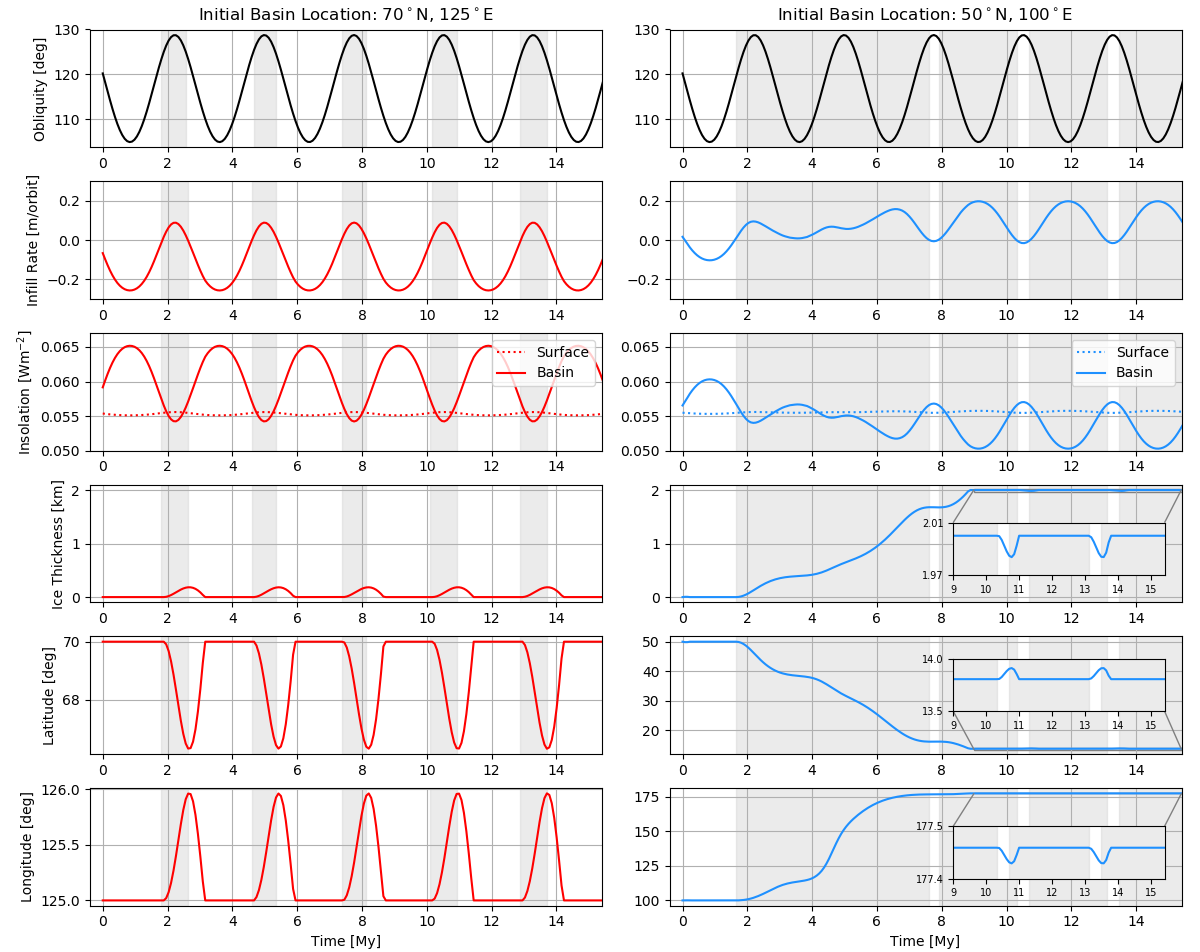}
		\caption{Time series of two example basins, highlighting the location and thickness excursions as a result of the obliquity cycle. The gray shading across all panels indicates time periods of positive infill rate. The red curves in the left panel correspond to the red path labeled 1 in Figure \ref{fig:paths}, and the blue curves in the right panel correspond to the blue path labeled 5. In the third panel from the top, the solid line is the orbit-averaged insolation onto the basin, and the dashed line is the orbit-averaged insolation on the remainder of the surface. The insets on the right side zoom in to show the excursions in more detail.}
		\label{fig:excursions}
	\end{center}
\end{figure*}

Figure \ref{fig:excursions} explains the time series behavior leading to latitude and ice thickness excursions for two examples basin paths: the paths labelled 1 (red, left panel) and 5 (blue, right panel) from Figure \ref{fig:paths}. The left panel shows an initially high-latitude basin, which never infills completely and instead oscillates back and forth between a lower latitude and its initial location. At its initial high latitude, the basin is receiving more insolation than the average of the rest of the surface, so the infill rate is negative. However, there is not yet any N$_2$ ice to sublime away, since the basin starts out empty, so the basin remains as is and no reorientation occurs. Eventually, when the obliquity is high, the basin insolation is lower than the rest of the surface, so the infill rate becomes positive and condensation occurs. This infill period is indicated by the gray shading. As the basin fills, the increase in mass causes an equatorward reorientation. Eventually, the combination of a lower latitude, thicker ice sheet, and lower obliquity cause the infill rate to drop below zero once again. This time, there is N$_2$ ice to sublime away ($\sim$180 m in this case), so the ice thickness decreases and the resulting reorientation returns the basin to its original location. This process repeats when the obliquity cycles back to high values.

Basins that infill with all available N$_2$, such as the blue path labeled 5 in Figure \ref{fig:paths}, also experience cyclic excursions from their final semi-stable location and ice thickness, as seen in the right panel and insets of Figure \ref{fig:excursions}. In this case, the initially mid-latitude basin begins the Infilling Phase after 2 My, when the obliquity is high enough, and eventually collects all of the available $N_2$ and reorients over another 6.5 My, reaching its final semi-stable latitude of 13.8\degree N and final ice thickness of 2 km. When the obliquity returns to high values, $>$127\degree, the basin receives slightly more insolation than the average of the rest of the surface so a short period of sublimation and poleward reorientation occurs. 16.8 m is ice is sublimed, which causes a 0.1\degreespace reorientation in latitude. However, the combination of increasing the latitude and decreasing the obliquity causes the infill rate to return to positive values and the ice sheet returns to the maximum 2 km thickness, returning the basin to its previous semi-stable location. 

These excursions have consequences for present-day Pluto. Currently, Pluto's obliquity is 120\degreespace and increasing. If SP follows the blue curve in Figure \ref{fig:excursions}, for example, then it could undergo a small excursion, reorienting a few tenths of a degree in latitude and losing a few tens of meters of N$_2$ ice in the next 0.4 My or so. This period of reorientation would last for 0.3 My. The previous excursion began around 2.4 My ago, and there could be geologic evidence of it on Pluto's present-day surface. \citet{Bertrand+2018} discuss the effect of N$_2$ condensation on geology in more detail. These excursions could also be responsible for refreshing the surface and helping to remove impact craters, along with glacial flow and convective overturning.

Table \ref{table:excursions} shows the amplitude of the location and thickness excursions for each of the paths.

\begin{table}[]
\caption{Amplitude of Location and Ice Thickness Excursions}
\label{table:excursions}
\resizebox{\textwidth}{!}{%
\begin{tabular}{c|c|c|c|}
\cline{2-4}
\multicolumn{1}{l|}{} & \textbf{Latitude Excursion} & \textbf{Longitude Excursion} & \textbf{Ice Thickness Excursion} \\ \cline{2-4} 
\multicolumn{1}{l|}{} & \textbf{\degree} & \textbf{\degree} & \textbf{m} \\ \hline
\multicolumn{1}{|c|}{\textbf{Path 1}} & -3.7 & 0.96 & 183 \\ \hline
\multicolumn{1}{|c|}{\textbf{Path 2}} & 0.1 & -0.09 & -44.9 \\ \hline
\multicolumn{1}{|c|}{\textbf{Path 3}} & 0.07 & -0.04 & -10.8 \\ \hline
\multicolumn{1}{|c|}{\textbf{Path 4}} & 0.02 & -0.002 & -2.3 \\ \hline
\multicolumn{1}{|c|}{\textbf{Path 5}} & 0.1 & -0.03 & -16.3 \\ \hline
\multicolumn{1}{|c|}{\textbf{Path 6}} & 0.07 & -0.03 & -10.8 \\ \hline
\end{tabular}%
}
\end{table}

\subsection{Effect of Initial Obliquity}\label{sec:obliquity}
All of the results shown above are initialized with Pluto's present-day obliquity of 120\degree. However, the initial obliquity at the time of ice sheet formation is unknown. Additionally, the period and magnitude of the obliquity cycle at the time of the ice sheet formation is also unknown, which is discussed more in Section \ref{sec:sma}. Here, we explore the effect of varying the initial obliquity on the final location of the basin, final thickness of the ice sheet, and the timing of the infill and subsequent excursions. 

\begin{figure*}
	\begin{center}
		\includegraphics[width =\textwidth]{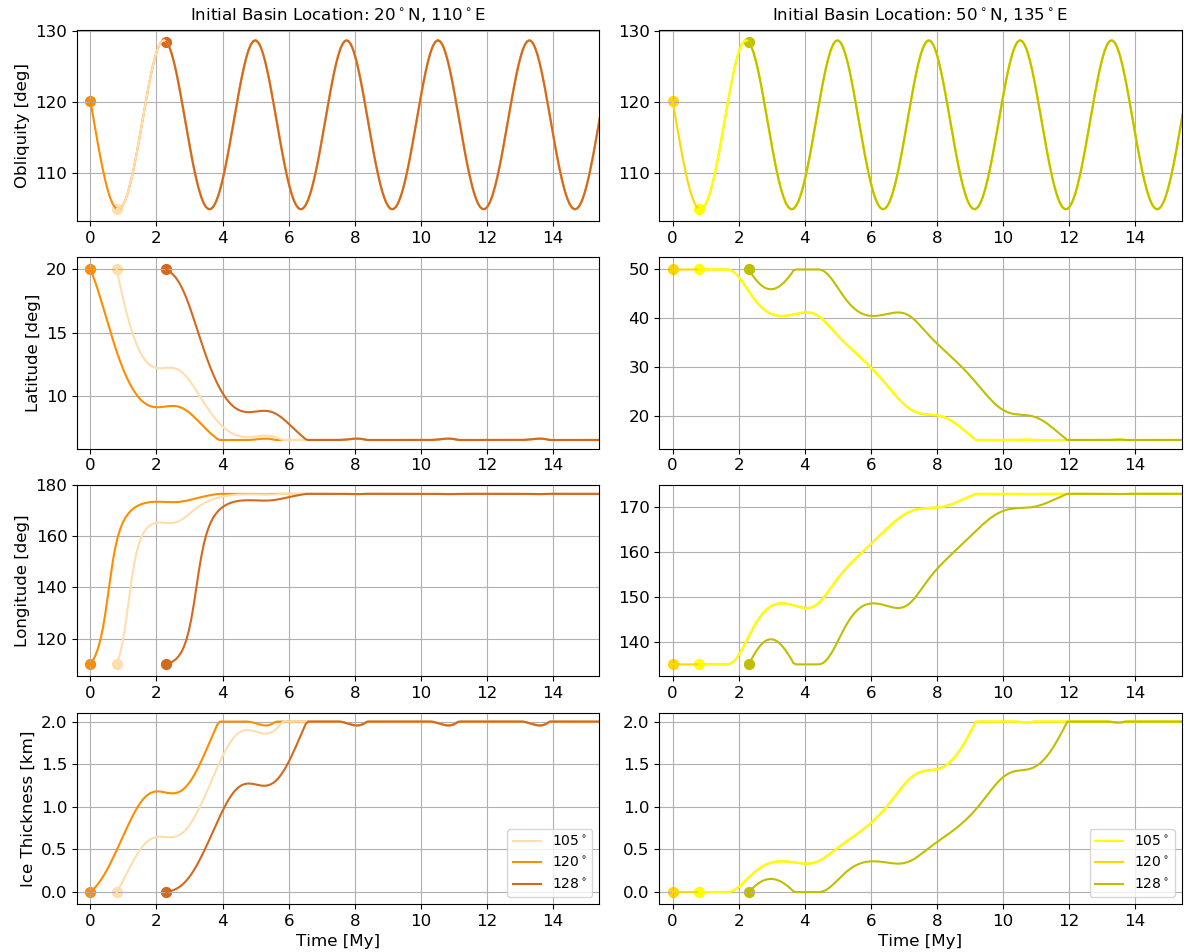}
		\caption{Time series of the obliquity, latitude and longitude of the basin, and thickness of the ice sheet for a 3-km basin with an 80 m GEL N$_2$ inventory beginning at 20\degree N, 110\degree E (left, corresponding to the orange path labeled 2 in Figure \ref{fig:paths}) and beginning at 50\degree N, 135\degree E (right, corresponding to the yellow path labeled 3 in Figure \ref{fig:paths}). The line shade indicates the initial obliquity: light orange and yellow have an initial obliquity of 105\degree, medium orange and yellow are 120\degree, and dark orange and yellow are 128\degree. Note that, in the right panel, the 105\degreespace curve lies directly on top of the 120\degreespace curve. The initial obliquity only affects the timing of events and not the final location, final ice thickness, nor the magnitude of the location or thickness excursions. }
		\label{fig:varyingobl}
	\end{center}
\end{figure*}

Figure \ref{fig:varyingobl} shows that changing the initial obliquity only affects the timing of events and not the final location, final ice thickness, nor the magnitude of the location or thickness excursions. It shows the location and ice thickness over time for a 3-km basin with an 80 m GEL N$_2$ inventory beginning at either 20\degree N, 110\degree E  or beginning at 50\degree N, 135\degree E with initial obliquities of 120\degreespace (the present-day value), 105\degreespace (the minimum of the obliquity cycle), and 128\degreespace (the maximum of the obliquity cycle). The 120\degreespace run starts at time t = 0, and the 105\degreespace and 128\degreespace runs begin 0.8 and 2.3 My later, respectively, when the phase of the obliquity cycle reaches the specified value. 

In all cases, the three line colors, representing different initial obliquities, all eventually converge. As seen in the left panel of Figure \ref{fig:varyingobl}, the different initial obliquity model runs converge to the same location and ice thickness after about 6 My for this initially low-latitude basin, while the initially intermediate-latitude basin, shown in the right panel, takes 12 My to converge. If in the 120\degreespace case, the basin enters the Infilling Phase before the obliquity cycle reaches 105\degreespace or 128\degreespace, then the 105\degreespace and 128\degreespace initial obliquity curves have to spend time ``catching up'' to the 120\degree case, delaying their infill and time to reach their final location. This is seen for both the 105\degreespace and 128\degreespace initial obliquity curves in the left panel, and the 128\degreespace initial obliquity curve in the right panel. However, since the initially-intermediate latitude basin in the right panel begins with a 2 My Stationary Phase period, in which it is waiting for a favorable obliquity for infill to begin, the 105\degreespace initial obliquity curve lies exactly on top of the 120\degreespace curve as soon as it begins (0.8 My after the 120\degreespace curve).

\subsection{Effect of Basin Deformation}\label{sec:deformation}
As the ice sheet grows, the mass of N$_2$ ice deforms the underlying crust, causing the ice sheet to sink lower into the surface. However, the magnitude for the deformation is dependent on assumptions about the interior structure of Pluto. Here, we investigate the effect of deformation by comparing our nominal  model to a model that ignores the effect of deformation entirely, equivalent to a perfectly rigid lithosphere with no elastic layer.

\begin{sidewaysfigure}
	\begin{center}
		\includegraphics[width =\textwidth]{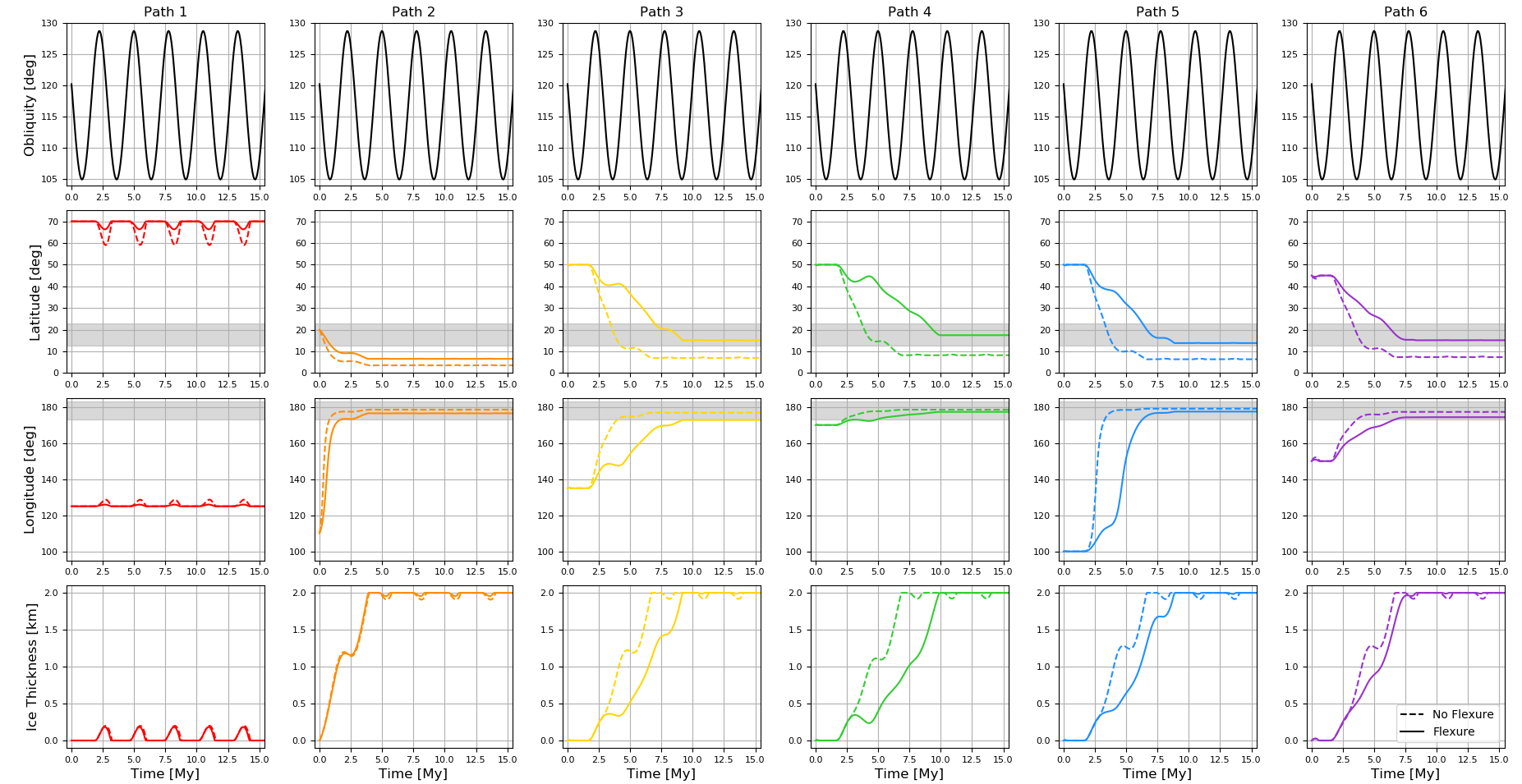}
		\caption{Comparison of the temporal behavior for Pluto's obliquity, the basin's latitude and longitude, and ice thickness for each path for the nominal model (including crustal deformation, solid line) and without crustal deformation (dashed line). For the latitude and longitude rows, the shaded region is $\pm$5\degreespace from the latitude or longitude of SP's present-day location. Obliquity is shown in black because it is the same for all paths; it is repeated in each column for ease of comparison. The color scheme is the same as that used in Figure \ref{fig:paths}.}
		\label{fig:noflex}  
	\end{center}
\end{sidewaysfigure}

Our model accounts for the deformation by varying the elevation of the basin floor (underneath the N$_2$ ice sheet) as a function of ice thickness. We solve for this deformation self consistently with the viscoelastic Love number interior structure model described in Section \ref{sec:model}. In short, for a 50 km thick elastic lithosphere, for every 1 km of N$_2$ ice in the basin, the basin floor sinks by 0.5 km. As the basin floor sinks, the top of the ice sheet sinks with it, so consequently the condensation surface is deeper and the infill rate is higher than it would be if the crust was not deformable. Figure \ref{fig:noflex} compares the temporal behavior of the model including crustal deformation with a variation of the model that does not allow the crust to deform. When crustal deformation is ignored, the basins reorient further in a shorter amount of time relative to when crustal deformation is included. As a consequence of crustal deformation, the present-day depth of the SP basin floor is deeper than the depth of the original, unfilled impact basin's floor. 

\subsection{Effect of Albedo}\label{sec:albedo}
In the preceding sections, the modeled surface, basin floor, and ice sheet all have an albedo of 0.75. However, the infilling rate $\dot{m}$ is directly proportional to $(1-A)$, so reducing the albedo increases the infilling rate, and vice versa. This section explores the effect this has on the final basin locations, depths, and ice thicknesses.

The principal effect of albedo is on timing. Increasing the albedo reduces the infilling rate, so the basins infill more slowly, all else being equal. With an albedo of zero, basins at all initial locations reached their final semi-stable locations within 6 My or so, while with an albedo of 0.75, many basins required 10 My to reach their final locations and enter the Cyclic Phase. 

\begin{figure}
	\begin{center}
		\includegraphics[width =\textwidth]{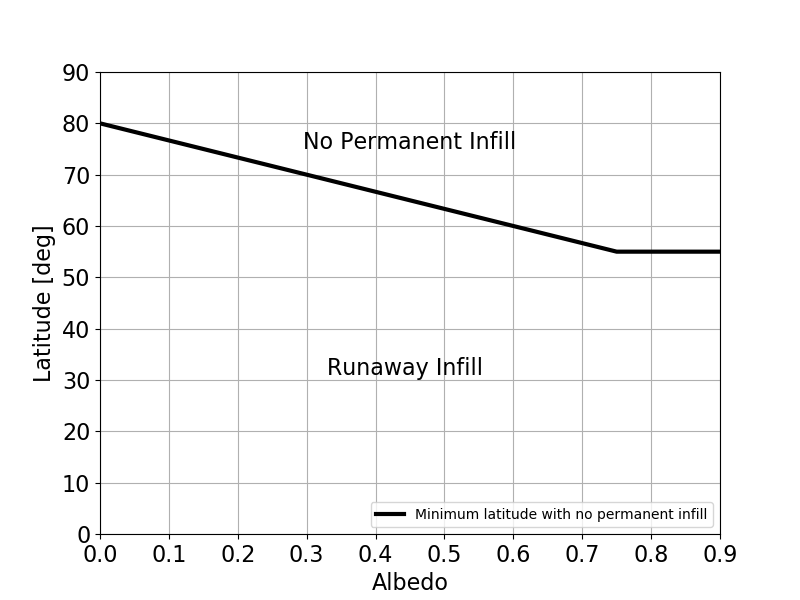}
		\caption{Dependence of the transition latitude between back-and-forth paths and runaway infill paths on albedo. Note that the latitude here is the lowest latitude at any longitude that exhibits back-and-forth reorientation; some basins at that latitude may exhibit runaway reorientation depending on their longitude.}
		\label{fig:transition}
	\end{center}
\end{figure}

Slower infill also has an indirect consequence on the initial locations that reorient to within $\pm$5\degreespace of the present-day location of SP. Initially high-latitude basins undergo only brief periods of positive infill rates, and when they accumulate enough N$_2$ ice during one of these periods, the subsequent reorientation to lower latitudes leads to a runaway process in which the basins infill completely and collect all available N$_2$. If the infill rate is lower, due to a higher albedo for example, it becomes more difficult to accumulate enough ice during the favorable-obliquity period, and the basins instead exhibit a back-and-forth reorientation where they never infill completely (in fact they never enter the Infilling Phase) and always return back to their initial location (e.g. the red path labeled 1 in Figure \ref{fig:paths}). Figure \ref{fig:transition} shows how the transition latitude between back-and-forth paths and runaway infill paths depends on albedo. The latitude decreases from 80\degreespace N at an albedo of zero to 55\degreespace for an albedo of 0.9. This means that, if the albedo is 0.9, no basins starting at or north of 55\degreespace will have final locations within $\pm$5\degreespace of the present-day location of SP. In terms of the regions defined in Figure \ref{fig:summary}, the lower boundary of Region 1 is defined by this the transition latitude, 55\degreespace for the A = 0.9 scenario. 

\begin{figure}
	\begin{center}
		\includegraphics[width =\textwidth]{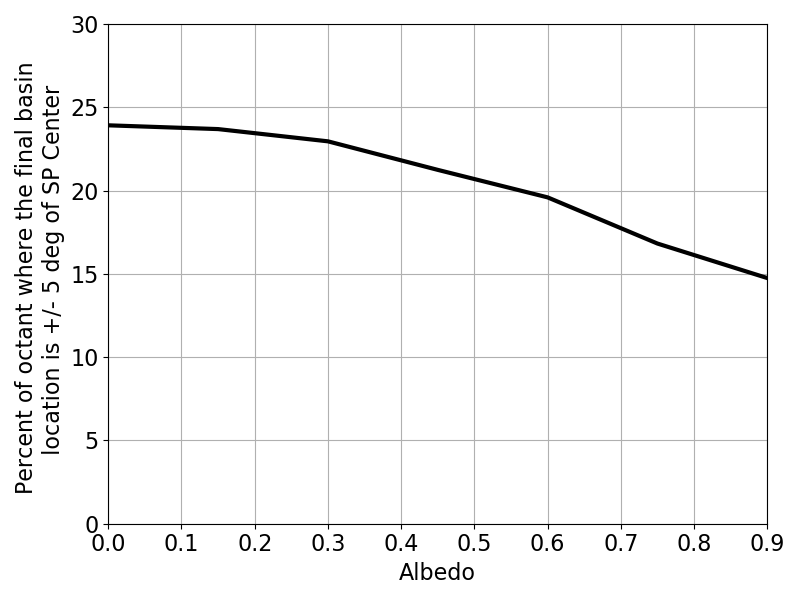}
		\caption{Fraction of the surface where the initial basin could be in order to reorient to within $\pm$5\degreespace of the present-day location of SP as a function of albedo. Note that this is the fraction relative to the region of interest (0\degree N to 90\degree N and 90\degree E to 180\degree E); to find the fraction relative to Pluto's entire globe, the values in this figure need to be divided by 8. }
		\label{fig:fracarea}
	\end{center}
\end{figure}

Figure \ref{fig:fracarea} shows the fraction of the surface where the initial basin could be in order to reorient to within $\pm$5\degreespace of the present-day location of SP. This is the fraction relative to the octant of interest (0\degree N to 90\degree N and 90\degree E to 180\degree E); to find the fraction relative to Pluto's entire globe, the values in Figure \ref{fig:fracarea} need to be divided by 8. Higher albedo surfaces have smaller initial areas that reorient to be within $\pm$5\degreespace of the present-day location of SP, in part due to the lower transition latitude between back-and-forth and runaway-infill reorientations discussed above. Figure \ref{fig:fracarea} shows that, if the basin was formed by an impact as suggested, impacts onto 15-25\% of the surface, mostly at intermediate latitudes, would be capable of reorienting to produce the observed present-day location of SP. 

\subsection{Effect of Pluto's Early Orbit} \label{sec:sma}
For the results presented above, we used Pluto's present-day orbital parameters: namely a semi-major axis of 40 AU and an eccentricity of 0.254. However, at the time of the ice sheet formation, which could be as long ago as 4 Gy, Pluto could have been in a much different orbit, potentially much closer to the Sun. For example, \citet{Nesvorny2015} modeled the orbital evolution of many Plutinos (other objects in the 3:2 resonance with Neptune) and found that Plutinos can spend significant time in orbits with semi-major axes less than 30 AU. As a simple test of how a reduction in semi-major axis affects our results, we repeated the analysis shown in Figure \ref{fig:success} for a 3-km deep basin and 80 m GEL N$_2$ inventory with a semi-major axis of 20 AU instead of 40 AU. We found that there were more initial basin locations which reorient to within $\pm$5\degreespace of SP's present-day location for the 20 AU semi-major axis orbit compared to the 40 AU case. Regions 3, 4, and 5 (see Figure \ref{fig:summary}) extended to higher latitude (70\degree N). Reducing the semi-major axis has a similar affect to reducing the albedo; both changes increase the absorbed insolation and therefore increase the infill rate.

\begin{figure*}
	\begin{center}
		\includegraphics[width =\textwidth]{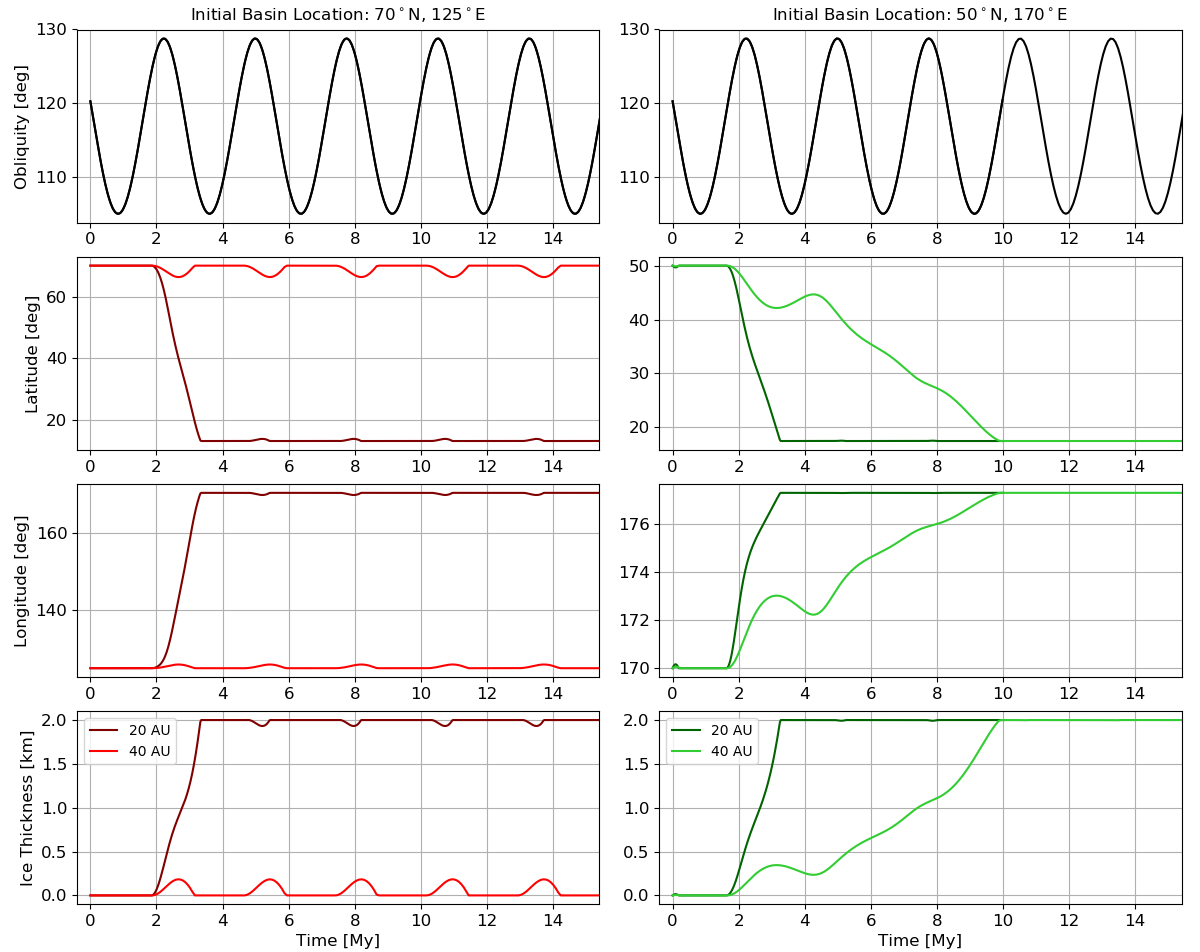}
		\caption{Time series of the obliquity, latitude and longitude of the basin, and thickness of the ice sheet for a 3-km basin with an 80 m GEL N$_2$ inventory beginning at 70\degree N, 125\degree E (left panel, corresponding to the red path labeled 1 in Figure \ref{fig:paths}) and 50\degree N, 170\degree E (right panel, corresponding to the green path labeled 4 in Figure \ref{fig:paths}). The line color indicates the orbit's semi-major axis: lighter colors are Pluto's present-day semi-major axis of 40 AU, and darker colors are a hypothetical past orbit's semi-major axis of 20 AU. }
		\label{fig:sma}
	\end{center}
\end{figure*}

Figure \ref{fig:sma} shows a comparison of results for a 20 AU and 40 AU orbit. In this figure, the basin is 3-km deep and there is 80 m GEL of available N$_2$ ice, and the basin's initial location is either 70\degree N, 125\degree E (left panel, red curves) or 50\degree N, 170\degree E (right panel, green curves). The differences between the 20 AU and 40 AU runs shown in the right panel are indicative of most initial basin locations. The final location and final ice thickness are not affected by reducing the semi-major axis in most cases, since a reduction in semi-major axis affects the $S_{surface}$ and $S_{basin}$ terms in Equation 3 proportionally. The surface pressure and temperature affect infill rates only through the latent heat, which changes slowly with temperature. The overall infill rate is quadrupled, so the 20 AU basins infill four times faster than the 40 AU basins. However, in some cases, the infill rate is high enough at 20 AU to allow complete infill, while at 40 AU the same basin never accumulated much ice and doesn't enter the Infilling Phase. The left panel of Figure \ref{fig:sma} shows an example of this, in which the 20 AU basin (dark red) forms a 2-km thick ice sheet and reorients close to the anti-Charon point, while the 40 AU basin (lighter red) never forms an ice sheet thicker than 0.25 km and cyclically reorients away from and back to its initial location. 

Figure \ref{fig:sma} assumes the same obliquity cycle as used throughout this work: a 2.7 My periodic cycle oscillating between 105\degreespace and 128\degree. However, since this cycle is determined by perturbations from other solar system bodies, the period and amplitude are likely to change if the orbit's semi-major axis is reduced from 40 AU to 20 AU. Additionally, the solar output from the Sun 4 Gy ago is potentially quite different from the present-day value. 
A more detailed analysis of a ``closer-in'' version of Pluto's orbit, accounting for these variables, is left for future work. 

\section{Conclusions} \label{sec:conclusions}
We present, for the first time, a coupled reorientation and climate model that can calculate the atmospheric condensation onto a ice sheet in a depression as a function of time, accounting for Pluto's varying obliquity and the possibility of true polar wander (TPW) as the mass contained within the basin grows. By stipulating that the final basin location must be within $\pm$5\degreespace of the present-day center of SP at 17.7\degree N, 178.2\degree E and that the top of the final ice sheet must be at an elevation of -2 km, we find that the final thickness of the ice sheet in the basin cannot be larger than 2 km, which is equivalent to a total mass of 1.4 x 10$^{18}$ kg or an 80 m global equivalent layer. More massive ice sheets cause too large of reorientations and the final basin ends up closer to the anti-Charon point than SP is observed to be. Based on the observed topography, the top of the ice sheet needs to be 2 km below the surrounding terrain, requiring that the initial basin depth be 2.5--3 km (final basin depth of 3--4 km) with a 1--2 km thick final ice sheet. The final ice thickness is subject to our assumption of the elastic thickness of the lithosphere; although as discussed earlier the effect is likely to be small due to the opposing effects of a larger remnant figure and less load compensation that result from thinner elastic lithospheres. With these requirements, our model finds that the basin most likely formed at an intermediate latitude, between 35\degree N and 50\degree N.
\begin{itemize}
    \item Mean Ice Thickness: 1--2 km
    \item Initial Basin Depth: 2.5--3 km
    \item Present-Day Basin Depth: 3--4 km
    \item Initial Location: intermediate latitudes ($\sim$ 35\degree N -- 50\degree N). 
\end{itemize}

The constraints presented in Section \ref{sec:results} and summarized above, 1--2 km for the maximum ice thickness and 3 or 4 km for the present-day depth of the basin, are both slightly smaller than previous estimates for those quantities. Convection models \citep{McKinnon+2016, Trowbridge+2016} imply an ice thickness of 3-10 km, but they are dependent on the unknown N$_2$ ice rheology. \citet{McKinnon+2016} estimate that ice sheets thicker than 1 km will undergo convection, so the ice sheets in this model would be able to support active convection. The depth of the basin itself is also not well known, besides needing to be 2 km deeper than the thickness of the ice sheet, to account for the observed elevation difference between the top of the ice sheet and the level of the average terrain. Earlier estimates placed the depth to the top of the ice sheet at 3--4 km (i.e. \citet{Moore+2016}), larger than the 2 km we use here. This is due in part to improvements to the global stereo topography model \citep{Schenk+2018}. Also, the -2 km elevation estimate is made relative to the average surface elevation, rather than to the topographically-high rim surrounding the basin. \citet{McKinnon+2016} argue that the depth of the original empty basin was likely less than 10 km, and the impact models from \citet{Johnson+2016} produce final compensated basins on the order of 5-10 km deep. Our assumptions about the initial gravity anomaly of the empty basin affects the final ice thickness and basin depth, and can be easily adjusted to reflect new information about Pluto's interior as it becomes available.

These constraints on the basin depth, ice thickness, and initial location are somewhat dependent on the assumptions made in the model, which are explained below.
\begin{enumerate}
    \item We assume that the empty basin has a net-zero gravity anomaly, implying that there is some form of compensation, like subsurface ocean uplift or an ejecta blanket \citep{Nimmo+2016,Keane+2016}. If the empty basin instead had a negative gravity anomaly (e.g. not fully compensated), the final ice thickness would be larger than 2 km. 
    \item We also make assumptions about Pluto's interior structure, which determines the  deformation in response to ice loading, and the rotational and tidal deformation preserved by a fossil figure. A weaker interior (e.g. due to a thinner elastic lithosphere or a smaller rigidity) results in more ice sheet compensation, which decreases TPW, and a smaller fossil figure, which increases TPW. These two effects nearly cancel each other, making the results presented here insensitive to the assumed interior structure.
    \item The topography of the basin floor, underneath the ice sheet, is unknown. We assume a flat bottom. If the basin floor instead is curved such that the ice sheet is thickest in the center and thinner at the edges, then our model's estimate of the ice thickness (where this is now measured at the center of the basin) would be slightly larger. For basins containing equal-mass ice sheets, curved-bottom basins will have deeper centers than flat-bottomed basins.
    \item We assume that Pluto reorients instantaneously in response to movements of volatiles across the surface. If Pluto does not respond this fast, then volatile transport may result in small-scale non-principal axis rotation, which would result in more complicated couplings between rotational dynamics and volatile transport.
\end{enumerate}

This TPW model accounts for the time variability of Pluto's climate and accounts for how reorientation changes the infill rate into the basin, unlike previous models. We find that most ice sheets form completely within 5--10 My, subject to the assumption of instantaneous reorientation after each timestep, but that this formation is not monotonic. The modelled basins do not infill at a constant rate from start to finish; instead the rate varies based on the current obliquity and the current latitude of the basin (which control the relative insolation onto the ice sheet) and also on the current depth to the top of the ice sheet. In some cases, the ice sheet begins to form, but then partially or completely sublimes away due to a change in obliquity. Even after reaching a final semi-stable state (e.g. collecting all of the available N$_2$ inventory), changes in the obliquity drive changes in the ice thickness and basin location every 2.7 My, on the order of 10 m in thickness and 0.1\degreespace in latitude. This cyclic TPW would continue to be active on present-day Pluto, with the next sublimation period expected to occur in 0.4 My, and the previous one ending 2 My ago. 

The origin of Pluto's Sputnik Planitia basin and ice sheet is a rich problem. This work adds to previously presented work \citep{Keane+2016, Nimmo+2016} that suggested that TPW is the mechanism that reoriented an impact basin to the present-day location of Sputnik Planitia. This work shows that the TPW hypothesis can and should be coupled with Pluto's climate and orbital evolution, and that this coupled model is also consistent with the observed features of Sputnik Planitia. Further understanding of the details of Pluto's orbit and climate early in the history of the solar system and how that affects the ice sheet's formation is warranted.

\acknowledgments
This work was supported by NASA ROSES SSW grant NNX15AH35G. A portion of this research was carried out at the Jet Propulsion Laboratory, California Institute of Technology, under a contract with the National Aeronautics and Space Administration (80NM0018D0004). A portion of this research was supported by the California Institute of Technology Joint Center for Planetary Astronomy postdoctoral fellowship. A portion of this work was supported by the National Aeronautics and Space Administration (80NSSC17K0724). Discussions with Pat McGovern helped frame the context. The authors would also like to express their appreciation of the New Horizons mission and its team for their exceptional work.


\bibliography{main}{}
\bibliographystyle{aasjournal}



\end{document}